\documentclass[useAMS,usenatbib,usegraphicx]{mn2e}

\usepackage{amsmath,amssymb,amsbsy,amsfonts}
\usepackage[british]{babel}
\usepackage{mdwmath}
\usepackage{color}
\usepackage{url}

\newcommand{\ud}{\mathrm{d}}
\newcommand{\sn}{\ensuremath{\mathrm{sn}}}
\newcommand{\cn}{\ensuremath{\mathrm{cn}}}

\newcommand{\s}[1]{{\text{\tiny $#1$ }}\hspace{-3pt}}
\newcommand{\avg}[1]{\left< #1 \right>}

\renewcommand{\theequation}{\thesection.\arabic{equation}}

\title[Relativistic accretion of a rotating cloud]{Analytic solutions to the
accretion of a rotating finite cloud towards a central object - II. Schwarzschild
spacetime}
\author[E. Tejeda, S. Mendoza \& J.C. Miller]
       {Emilio Tejeda$^1$\footnotemark[1], Sergio Mendoza$^2$\footnotemark[1] 
	and John C. Miller$^{1,3}$\thanks{E-mail: tejeda@sissa.it (ET), 
	sergio@astro.unam.mx (SM), miller@sissa.it (JCM).} \\
        $^1$ SISSA \& INFN, Via Bonomea 265, 34136, Trieste, Italy \\
	$^2$ Instituto de Astronom\'ia, Universidad Nacional
        Aut\'onoma de M\'exico, AP 70-264, Distrito Federal 04510, Mexico\\
	$^3$ Department of Physics (Astrophysics), 
	University of Oxford, Keble Road, Oxford OX1 3RH, UK }

\pagerange{\pageref{firstpage}--\pageref{lastpage}}

\begin{document}

\maketitle

\label{firstpage}

\begin{abstract} 
 We construct a general relativistic model for the accretion flow of a rotating 
finite cloud of non-interacting particles infalling onto a Schwarzschild black 
hole. The streamlines start at a spherical shell, where boundary conditions are 
fixed with wide flexibility, and are followed down to the point at which they either
cross the black hole horizon or become incorporated into an equatorial thin disc.
Analytic expressions for the streamlines and the velocity field are given, in terms
of Jacobi elliptic functions, under the assumptions of stationarity and ballistic
motion. A novel approach allows us to describe all of the possible types of orbit
with a single formula. A simple numerical scheme is presented for calculating the
density field. This model is the relativistic generalisation of the Newtonian one
developed by \citet{mendoza} and, due to its analytic nature, it can be useful in
providing a benchmark for general relativistic hydrodynamical codes and for exploring
the parameter space in applications involving accretion onto black holes when the
approximations of steady state and ballistic motion are reasonable ones.
\end{abstract}


\section{Introduction}
\label{intro}

For several decades, accretion onto black holes has been recognised as the mechanism
behind some of the most powerful astrophysical phenomena \citep[see 
e.g.][]{frank02}. It has been extensively studied in the context of Active 
Galactic Nuclei (AGN) \citep{genzel10}, Gamma-Ray Bursts (GRBs) \citep{piran04},
X-ray binaries \citep{king95}, compact binary coalescence \citep{hughes09} and tidal
disruption of stars by black holes \citep{rosswog09}.

Spherical accretion onto a black hole was found to have low efficiency for converting
gravitational potential energy into emitted radiation \citep{shapiro73} and so
rotation of the accreting matter has been invoked to give the increased efficiencies
required by observations. Rotation in the flow often leads to a centrifugal hang-up
with the formation of an accretion disc around the central object through which
material inspirals slowly under the action of dissipative mechanisms which can give
efficient conversion of gravitational potential energy into radiation. The studies of
accretion discs involve many physical inputs including magnetohydrodynamic
turbulence, the behaviour of highly ionised and degenerate matter, radiative
processes and radiation transfer, the formation of strong shocks, nuclear burning,
etc; however, gravity and rotation play the dominant role in determining the overall
accretion regime and bulk dynamics.

Modern accretion theory began with the pioneering work by \citet{bondi52}, in 
which he gave the analytical solution for the steady spherical accretion of an 
ideal gas cloud onto a central object in Newtonian gravity. The relativistic
extension of this was then developed by \citet{michel72} who took a Schwarzschild
black hole as the central accretor. Rotating inflows were discussed by \citet{pb68},
who introduced the idea of accretion discs which was then developed further by
\citet{shakura73} and \citet{novikov73}.

Formation of an accretion disc by infall of a rotating gas cloud onto a central 
object was first treated by \citet{ulrich} in the context of star formation and 
accretion discs around protostars. In that Newtonian analysis, the accreting gas
was considered to start falling in from a spherical shell located infinitely far
away from the central object where it had uniform density. Furthermore, it was
assumed that the shell was rotating uniformly and that the fluid elements were
following parabolic trajectories. The disc forms in the plane perpendicular to
the cloud's rotation axis as a result of the collision of streamlines coming
from mirror-symmetric points on opposite sides of the equator. \citep[See
e.g.][for a clear description of Ulrich's model and the disc formation
process.]{cassen81, lin90, stahler94, mendoza04, nagel07} The relativistic
extension of Ulrich's model was investigated numerically by \cite{belo01}, in
relation to wind-fed X-ray binaries, and has also been studied analytically by
\cite{huerta07} (HM07 hereafter) and \citet{mendoza08}. In the paper to which
the present one is a follow-up \citep[][- referred to as Paper I in the
following]{mendoza}, Ulrich's model was extended, within Newtonian gravity, by
considering a finite cloud radius and allowing for more general boundary
conditions.

In the present paper, we give a full relativistic generalisation of Paper I, 
studying the stationary inflow of a rotating cloud of non-interacting particles 
around a Schwarzschild black hole. We shall assume that the particles follow
ballistic trajectories and thus, that flow lines correspond to timelike geodesics of
Schwarzschild spacetime. Building on previous analytical studies \citep[see
e.g.][]{chandra}, we introduce a novel approach which allows us to describe all of
the different types of trajectory with a single analytic expression. The model tracks
the infall of particles injected from a finite spherical surface centred on the
accretor where the density and velocity distribution of the particles are fixed with
wide flexibility. Due to the rotation, a disc-like structure forms in the equatorial
plane. In the present work we focus on the accretion flow feeding the disc, and not
on the behaviour of matter within the disc itself, where the ballistic approximation
certainly breaks down. Our aim is to give an analytic description of this idealised
accretion flow, starting far away from the central object and following it down to
the vicinity of the equatorial disc, where we then stop our calculations. In the
present work, the disc and the black hole are treated just as passive sinks of
particles and energy.

In this paper we follow an analytic approach similar to that in HM07, although in
Section~\ref{ulrich} we show that the final expressions of HM07 need to be modified.
Furthermore, in \cite{belo01} and HM07 the authors considered as the boundary
condition a uniform-density, rigidly-rotating dust shell with all of the fluid
elements following parabolic-like motion. In contrast, the present model allows for
more general distributions of density, rotation profile, accretion rate and particle
energy.

We propose to use this analytic calculation as a benchmark to test the ability of
general relativistic hydrodynamical codes in recovering geodesic motion when weakly
interacting particles moving on a fixed background metric are considered.

Moreover, the model presented here might also constitute a valuable tool for
exploring the effect of different velocity and density distributions on the 
overall accretion scenario, in astrophysical applications where the ballistic 
and steady state conditions are reasonable approximations such as supersonic 
accretion from a wind-fed X-ray binary, as in \cite{belo01}; sub-Eddington gas 
accretion onto galactic nuclei \citep[see e.g.][]{blaes07}; or collapsing 
stellar cores, as in \cite{lee} (LR06 hereafter) \citep[see also][]{lopezcamara09,
taylor11}. Exploring the role of different boundary conditions, before full
hydrodynamical simulations are performed, becomes especially useful if one considers
that in most of these situations the angular momentum distribution of the accreting
matter is highly uncertain, even if it is clear that rotation does play a crucial
role in determining the overall accretion efficiency.

As an example of the validity of such an exercise, in Section \ref{example} we 
compare the velocity and density fields, as predicted by the analytic model, with
ones from a roughly equivalent numerical simulation from LR06. In that work, the
authors explored numerically the accretion flow of the collapsing interior of a
massive star towards a newborn black hole, mimicking relativistic effects by means of
a \cite{paczyski80} pseudo-Newtonian potential.

 The structure of our paper is as follows. In Section~\ref{model} we present the
model and its assumptions. In Section~\ref{streamlines}, a general expression for the
fluid streamlines is given. In Section~\ref{velocity} the velocity fields are
described as seen by observers located at infinity and by local observers. Using the
continuity equation, a numerical scheme for calculating the density field is
developed in Section~\ref{density}. The model is illustrated with a simple choice of
boundary conditions and then compared against a numerical simulation from LR06 in
Section~\ref{example}. In Section~\ref{ulrich}, we adopt instead the boundary
conditions of the \cite{ulrich} model, and give its general relativistic extension.
Finally, in Section~\ref{newton}, the non-relativistic limit is considered and 
the results given in Paper I are recovered.

\setcounter{equation}{0}
\section{Model description}
\label{model}

  We are interested here in modelling a rotating cloud of particles falling
towards a central black hole with mass $M$, whose exterior gravitational field
is described by the Schwarzschild metric. We assume that the accretion flow has
reached a stationary situation characterised by a constant accretion rate
$\dot{M}$, where from now on, a dot denotes a derivative with respect to the
proper time $\tau$. Additionally, we assume that the gravitational field of the
black hole is the main factor determining the fluid dynamics. We therefore
neglect the effects of fluid self gravity, pressure gradients, fluid viscosity,
radiation pressure, neutrino capture, etc. In other words, we give a ballistic
treatment of the fluid flow.

\begin{figure}
\begin{center}
  \includegraphics[scale=0.5]{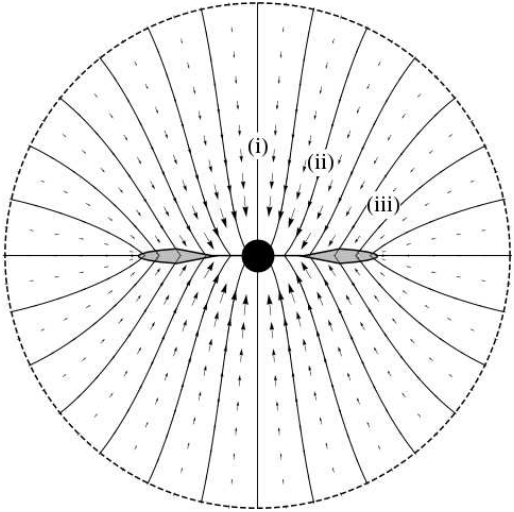}
\end{center}
\caption{Schematic illustration of the present model. Boundary conditions are fixed
at a shell with radius $r_\s{0}$ and fluid elements then move along ballistic
trajectories. We assume an axisymmetric distribution of angular momentum with its
magnitude increasing monotonically towards the equator. The streamlines can be
divided into three groups: (i) ones with low angular momentum that accrete directly
into the black hole; (ii) ones with intermediate angular momentum which arrive at the
equator but do not find a circularisation radius; (iii) ones with large enough
angular momentum to be incorporated into an equatorial Keplerian-type accretion
disc.}
\label{f1}
\end{figure}

Consider a cloud subdivided into equal mass fluid elements. We let $\varrho$ be 
the matter density and $n$ be the baryon number density, and introduce an 
average baryonic rest mass, $\avg{m}$, such that $\varrho=\avg{m}n$. We assume 
that fluid elements are continuously injected from a spherical shell of radius 
$r_\s{0}$. This shell represents the outer boundary of the model where the 
fluid properties are fixed as

\begin{gather}
\varrho_\s{0} =
\varrho\left(r_\s{0},\,\theta_\s{0},\,\phi_\s{0}\right),\label{e0.1}\\
\dot{r}_\s{0} = \dot{r}\left(r_\s{0},\,\theta_\s{0},\,\phi_\s{0}\right),
\label{e0.2}\\
\dot{\theta}_\s{0} =
\dot{\theta}\left(r_\s{0},\,\theta_\s{0},\,\phi_\s{0}\right), \label{e0.3}\\
\dot{\phi}_\s{0}
=\dot{\phi}\left(r_\s{0},\,\theta_\s{0},\,\phi_\s{0}\right), \label{e0.4}
\end{gather}

\noindent with $\dot{r}$, $\dot{\theta}$ and $\dot{\phi}$ being the radial, 
polar and azimuthal velocity components, respectively. Figure~\ref{f1} shows a 
schematic diagram of the accretion scenario.

We take the four distribution functions in eqs.\,\eqref{e0.1}-\eqref{e0.4} to be
differentiable and to be symmetric with respect the equatorial plane, i.e. 

\begin{equation}
  f(\theta_\s{0}) = f(\pi-\theta_\s{0}),
\label{e0.5}
\end{equation}

\noindent with $f=\varrho_\s{0},\, \dot r_\s{0},\, \dot\theta_\s{0},\,
\dot\phi_\s{0}$.

Using the boundary distributions for $\dot r_\s{0}$ and $\varrho_\s{0}$, one
calculates the total accretion rate as

\begin{equation}
\dot M = -r_\s{0}^2 \int\limits_0^{2\pi}\int\limits_0^\pi \varrho_\s{0}\,\dot
r_\s{0}\,\sin\theta_\s{0}\,\ud\theta_\s{0}\,\ud\phi_\s{0}.
\label{e0.6}
\end{equation}

Given the ballistic approximation, we have that the specific energy, $E$, and
the specific angular momentum, $h$, are conserved along streamlines. Clearly
their actual values vary from line to line and will be distributed in a certain 
fashion (see eqs.\,\ref{e2.10} and \ref{e2.11}) as a function of the boundary 
conditions in eqs.\,\eqref{e0.1}-\eqref{e0.4}.
 
It is a feature of the Schwarzschild spacetime that, for a given mass of the 
central object and a given streamline, with energy $E(\theta_\s{0},\,
\phi_\s{0})$, there exists a critical value of the specific angular
momentum,\footnote{Note that taking $h_c=2\,r_sc$ as the critical value, is
valid only for streamlines with parabolic-like energies, i.e. $E=c^2$.} $h_c$,
such that if $h(\theta_\s{0},\,\phi_\s{0})<h_c$ then the corresponding fluid
line crosses the black hole horizon (located at $r_s= 2\,GM/c^2$) before
reaching the equatorial plane. On the other hand, for
$h(\theta_\s{0},\,\phi_\s{0})>h_c$ the streamline reaches the equator and
encounters there an equivalent streamline coming from $(\pi - \theta_\s{0},\,
\phi_\s{0})$. In a real physical situation one expects that if these collide
supersonically then two shock fronts will appear above and below the equator,
with the fluid being incorporated into a disc-like structure. Provided that
there is an efficient dissipation mechanism, the shock fronts will remain pinned
down to the equator and the disc will remain thin. It is clear that the study of
the disc dynamics requires a full hydrodynamical treatment, in which
redistribution of angular momentum and energy losses are self consistently taken
into account, but such an analysis lies beyond the scope of the present work.
Instead, we shall just assume here that an efficient mechanism dissipates all of
the kinetic energy associated with the vertical component of the velocity, in
such a way that an infinitesimally thin disc forms in the equatorial plane which
is then taken to act as a passive energy sink. See \citet{lopezcamara09} for a
full hydrodynamical simulation of a collapsar in which they show that an
isothermal disc does indeed remain thin.

In principle one could relax the condition in eq.\,\eqref{e0.5} and not assume 
any particular symmetry for the fluid properties at the boundary. However, in 
that case we would not have the symmetric collision of streamlines described 
above. This would lead to formation of a warped disc, making the situation much 
more complicated.

Following LR06, the streamlines can be divided into three groups depending on 
the value of their specific angular momentum (see Figure~\ref{f1}): (i) 
streamlines with low angular momentum, which go directly into the black hole; 
(ii) streamlines with intermediate angular momentum, which form a small disc
within which accretion proceeds on a free-fall time scale \cite[this situation
corresponds to the one explored numerically by][]{belo01}; (iii) streamlines 
with larger angular momentum, which have sufficient centrifugal support on their
arrival at the equator so that subsequent accretion would occur on a viscous
time scale in a Keplerian-type accretion disc.

Now consider the situation in which two neighbouring streamlines start 
approaching each other in such a way that they would intersect. This type of 
encounter is qualitatively different from the head-on collision described 
above, since here, and with a full hydrodynamical treatment, the two approaching
streamlines would be prevented from intersecting by the smooth action of
pressure forces. It is clear, however, that this cannot be handled within the
ballistic approximation and so we must restrict our analysis to distribution
functions for which streamlines do not cross before reaching the equator.

Provided that there are no early intersections, then we can use the initial 
angular position, $(\theta_\s{0},\,\phi_\s{0})$, as a label of the individual 
streamlines. In the next Section we give an analytic expression for the 
streamlines that, for any given radius $r$, constitutes a mapping
$(\theta_\s{0},\, \phi_\s{0}) \leftrightarrow (\theta,\,\phi)$. The condition
of no intersection is equivalent to requiring that the Jacobian of this 
transformation should be non-singular, i.e.

\begin{equation}
J = \left(\frac{\partial\theta}{\partial\theta_\s{0}}\right)
 \left(\frac{\partial\phi}{\partial\phi_\s{0}}\right) -
 \left(\frac{\partial\theta}{\partial\phi_\s{0}}\right)
 \left(\frac{\partial\phi}{\partial\theta_\s{0}}\right) > 0.
\label{e0.8}
\end{equation}

\noindent While formal, the actual evaluation of this for given boundary 
conditions is not a trivial task in general. However, if there is axisymmetry it is a
sufficient condition that the specific angular momentum should be a non-decreasing
function of $\theta_\s{0}$ going from the polar axis to the equator, and that its
magnitude should be sufficiently small so that no fluid elements reach a turning
point in their trajectories before crossing the equator. For more general cases, it
seems best to proceed by trial and error. 

\setcounter{equation}{0}
\section{Streamlines}
\label{streamlines}

  Under the ballistic approximation, the streamlines of the accretion model
correspond to trajectories of freely-falling test particles (i.e. timelike
geodesics) in a Schwarzschild spacetime. In this Section we give a general
analytic expression for the spatial projection of these geodesic lines.

\subsection{Description in the orbital plane}

Let us consider a given fluid element starting to fall in from the point
$(r_\s{0}, \theta_\s{0}, \phi_\s{0})$. We denote by $\mathcal{O}$ the general
frame of reference with coordinates $(t, r, \theta, \phi)$ having its polar axis
coinciding with the rotation axis of the cloud.

In a general spacetime, the geodesic equations consist of a second order, non-linear,
coupled system of four differential equations.  Nevertheless, with the aid of the
underlying symmetries of the Schwarzschild spacetime, it is found that the
corresponding geodesic equations become separable and reduce to a set of four first
order differential equations \citep[see e.g.][]{novi}. A further simplification
coming from the spherical symmetry is that the particle motion is confined to a
single plane. We denote by $\mathcal{O}'$ a frame of reference with coordinates $(t,
r, \vartheta, \varphi)$, especially adapted such that its equatorial plane
($\vartheta=\pi/2$) coincides with the orbital plane.

  From the view point of $\mathcal{O}'$, the geodesic equation associated with 
the polar angle becomes $\dot\vartheta=0$. This confirms that the whole 
trajectory stays in a single plane. On the other hand, the equations
corresponding to the time and azimuthal coordinates lead to two first integrals 
of motion, namely the relativistic total specific energy, $E$, and the specific 
angular momentum, $h$, defined as:

\begin{gather}
h= r^2\dot{\varphi},\label{e1.2}\\
E= \left(1-\frac{r_s}{r}\right)c^2 \dot{t}.
\label{e1.3}
\end{gather}

  With the aid of eqs.\,\eqref{e1.2} and \eqref{e1.3} together with the
normalisation condition for the 4-velocity\footnote{Throughout this paper we
use Einstein's summation convention with Greek indices running over spacetime
coordinates while Latin indices run just over the spatial components.}
($u^\s{\mu}=\ud x^\s{\mu}/ \ud \tau$), $u^\s{\mu}u_\s{\mu}=-c^2$, one gets the
following equation governing the proper time evolution of $r$ for the fluid 
element

\begin{equation}
\left(\frac{\ud r}{\ud\tau}\right)^2 = \varepsilon
  +  \frac{2\, GM}{r} -\frac{h^2}{r^2} + \frac{r_s h^2}{r^3},
\label{e1.4}
\end{equation} 

\noindent where we have introduced the re-scaled energy

\begin{equation}
\varepsilon=\frac{E^2 - c^4}{c^2}.
\label{e1.5}
\end{equation}

\noindent This definition of energy is convenient because it facilitates 
comparison with the Newtonian case since, in the non-relativistic limit, 
$\varepsilon$ converges to twice the Newtonian total energy.

  Since we are interested in a stationary regime, it is convenient to recast 
the time derivative in eq.\,\eqref{e1.4} as a derivative with respect to the 
azimuthal angle $\varphi$ ($ \mathrm{d} / \mathrm{d} \tau = \dot{\varphi } \, 
\mathrm{d} / \mathrm{d} \varphi$). Doing this together with the aid of 
eq.\,\eqref{e1.2} allows us to rewrite eq.\,\eqref{e1.4} as

\begin{equation}
  \frac{\ud r }{\ud \varphi } = -\frac{\sqrt{\mathcal{R}(r)}}{h},
\label{e1.6}
\end{equation}

\noindent with

\begin{equation}
  \mathcal{R}(r) =  \varepsilon\,r^4 + 2\,GM r^3 - h^2\left( r^2 - r_s\,r
\right).
\label{e1.7}
\end{equation}

\noindent The minus sign in eq.\,\eqref{e1.6} is needed because, in the current
accretion scenario, $\varphi$ increases while $r$ decreases as the fluid
elements approach the equator.

Being a fourth degree polynomial, \( \mathcal{R}(r) \) has four roots (with
possible multiplicity) and we can write it as

\begin{equation}
\mathcal{R}(r) = \varepsilon(r-r_\s{a})(r-r_\s{b})(r-r_\s{c})(r-r_\s{d}).
\label{e1.8}
\end{equation}

\noindent Explicit expressions for the roots are given in the Appendix. It is clear
that $r = 0$ is one of them; we treat it in the same way as the others though, since 
keeping an explicit reference to it allows us to give a general expression for the
streamlines. Given that one root is zero, then necessarily at least one other root
must be real. The remaining two roots can either also both be real, or can form a
complex conjugate pair.

 After some analysis of \( \mathcal{R}(r) \) \citep[see, e.g.][]{mielnik} it 
follows that if $\varepsilon<0$, all of its real roots are non-negative, while 
for $\varepsilon>0$, there is exactly one negative root. For $\varepsilon=0$, 
one of the roots has diverged to infinity and $\mathcal{R}(r)$ reduces to a 
third order polynomial. This case is treated in further detail in 
Section~\ref{ulrich}.

From eq.\,\eqref{e1.6} it is clear that the radial motion is restricted by the 
condition $\mathcal{R}(r)>0$ and so $r$ is either bracketed in between two 
consecutive real roots\footnote{Whenever these bracketing roots are positive, 
they correspond to turning points for which the radial velocity vanishes and 
the particle goes from moving inwards to moving outwards, or vice versa.} 
(bounded motion) or is unbounded above.

Let $r_\s{a}$ be the lower bound (periastron). (This includes the possibility 
$r_\s{a}=0$, which corresponds to a plunge orbit.) 

If the four roots are real and the motion is upper bounded, we let $r_\s{b}$ be 
the upper bound (apastron), while for unbounded motion, we let $r_\s{b}$ be the 
only negative root. We then take the two remaining roots, $r_\s{c}$ and 
$r_\s{d}$, such that $r_\s{c}\le r_\s{d}$.

If two of the roots are complex, we call those roots $r_\s{b}$ and $r_\s{c}$, 
with $r_\s{b}=r_\s{c}^*$, and let $r_\s{d}$ be the remaining real root. This 
case, with complex roots, represents the purely-relativistic ``capture'' type 
of orbit, for which particles with non-vanishing angular momentum do not
have strong enough centrifugal support to prevent them from falling directly
into the black hole.

We set the origin of $\varphi$ at the periastron of the orbit, i.e., 
$r(\varphi=0) = r_\s{a} $. The solution of eq.\,\eqref{e1.6} is then equivalent 
to finding solutions to the following quadrature problem:

\begin{equation}
  \int^{r }_{r_\s{a} } \frac{\ud r' } {\sqrt{\mathcal{R}(r')}}=
    -\frac{\varphi}{h} .
\label{e1.9}
\end{equation}

\noindent This integral is solvable in terms of Jacobi elliptic functions, as has 
been discussed several times in the literature \citep[see, e.g.][]{hagihara,
darwin59, mielnik, metzner, chandra, miro}. These works have followed several 
different approaches, focusing on different aspects of the solution and using
different notations. Labelling the roots in the way described above enables us to
give a novel description of the different types of trajectory by means of a single
analytical expression, which summarises the results found in the previous work:

\begin{equation}
r = \frac{ r_{\s{b}} 
( r_{\s{d}} - r_{\s{a}} ) - r_{\s{d}} 
( r_{\s{b}} - r_{\s{a}} ) \cn^2 (\xi,\,k) }
{ r_{\s{d}} - r_{\s{a}} 
- ( r_{\s{b}} - r_{\s{a}} ) \cn^2 (\xi,\,k) },
\label{e1.10}
\end{equation}

\noindent with

\begin{equation}
\xi=\frac{\sqrt{\varepsilon(r_\s{a}-r_\s{c})(r_\s{d}-r_\s{b})}}{2\,h}\,\varphi,
\label{e1.11}
\end{equation}

\noindent where $\cn \left(\xi,\,k\right)$ is a Jacobi elliptic function with
modulus $k$ given by

\begin{equation}
k  = \sqrt{\frac{(r_\s{b} -r_\s{a} )(r_\s{d}-r_\s{c})}
{(r_\s{d} -r_\s{b} )(r_\s{c}-r_\s{a} )}}.
\label{e1.12}
\end{equation}

Note that, though general, the expressions in eqs.\,\eqref{e1.10}-\eqref{e1.12} 
should be handled with care in two particular cases: when complex roots are 
involved and when $\varepsilon\rightarrow0$. In the first case, some
intermediate terms will be complex quantities even though the final result will 
always be a real number. In the Appendix we give alternative expressions for 
this first case, while the second case is discussed in Section~\ref{ulrich}.

\subsection{Relation between the frames of reference}
\label{angles}

In this subsection we relate the descriptions made in $\mathcal{O}$ and in 
$\mathcal{O}'$, noting that, for fixed $t$ and $r$, the Schwarzschild metric 
\citep[see, e.g.][]{MTW} reduces to that of an ordinary 3-sphere and, 
hence, basic rotation operations can be performed. The easiest way to relate 
$\varphi$ to the angles $\theta$ and $\phi$ is by means of introducing the 
turning point in the polar motion, i.e., a point $\theta_\s{a}$ for which

\begin{equation}
 \dot \theta (\theta_\s{a}) = 0.
\label{e2.1} 
\end{equation}
 The polar motion of a non-equatorial trajectory is always characterised by two 
turning points which are symmetric with respect to the equator. That is, if 
$\theta_\s{a}$ satisfies eq.\,\eqref{e2.1} then $\pi-\theta_\s{a}$ does as 
well. Here we choose $\theta_\s{a}$ in the same hemisphere as $\theta_\s{0}$.
 
Noting that the polar axis in $\mathcal{O}'$ is tilted with respect to the one 
in $\mathcal{O}$ by an angle $\pi/2 -\theta_\s{a}$ and performing a series of
rotation operations,\footnote{First make a rotation of $-\varphi_\s{a}$ about
the $z$ axis in the $\mathcal{O}'$ frame, then one of $\pi/2 -\theta_\s{a}$
about the resulting $y$ axis, followed by one of $\phi_\s{a}$ about the final
$z$ axis.} one obtains the following relationships

\begin{gather}
\cos (\varphi-\varphi_\s{a}) = \frac{\cos \theta}{\cos \theta_\s{a} }, 
\label{e2.2} \\ 
\cos (\phi-\phi_\s{a})  = \frac{\cot \theta }{\cot \theta_\s{a} }.
\label{e2.3}
\end{gather}

 Differentiation of eqs.\,\eqref{e2.2} and \eqref{e2.3} with respect to 
the proper time gives

\begin{gather}
 \dot \varphi = \frac{\sin\theta\,\dot
\theta}{\sqrt{\cos^2\theta_\s{a}-\cos^2\theta}},
\label{e2.4}\\
 \sin\theta\,\dot \phi = \frac{\sin\theta_\s{a}\,\dot
\theta}{\sqrt{\cos^2\theta_\s{a}-\cos^2\theta}}.
\label{e2.5}
\end{gather}

\noindent Evaluating eq.\,\eqref{e2.5} at $\theta_\s{0}$ gives the following 
expression for $\theta_\s{a}$ in terms of known quantities

\begin{equation}
 \sin\theta_\s{a} = 
 \frac{\sin^2\theta_\s{0}\,\dot\phi_\s{0}}
{\sqrt{\dot\theta_\s{0}^2+\sin^2\theta_\s{0}\,\dot\phi_\s{0}^2}} .
\label{e2.6}
\end{equation}

\noindent From this last equation it is clear that if $\dot\theta_\s{0}=0$ then
$\theta_\s{a}=\theta_\s{0}$.

On the other hand, evaluation of eqs.\,\eqref{e2.2} and \eqref{e2.3} at 
$(\theta_\s{0},\,\phi_\s{0})$ allows us to calculate $\varphi_\s{a}$ and
$\phi_\s{a}$ through

\begin{gather}
\varphi_\s{a}  = \varphi_\s{0}  - 
\cos^{-1}\left(\frac{\cos\theta_\s{0}}{\cos\theta_\s{a}} \right) ,
\label{e2.7}\\
\phi_\s{a}  = \phi_\s{0} -
\cos^{-1}\left(\frac{\cot\theta_\s{0}}{\cot\theta_\s{a}} \right) ,
\label{e2.8}
\end{gather}

\noindent where, from eqs.\,\eqref{e1.10} and \eqref{e1.11},

\begin{equation}
\begin{split}
 \varphi_\s{0} = & \frac{-2\,h}
{\sqrt{\varepsilon (r_\s{a}- r_\s{c})(r_\s{d}-r_\s{b})}} \times \\
&\qquad \cn^{-1}\left(\sqrt{\frac{(r_\s{d}-r_\s{a})(r_\s{b}-r_\s{0})}
{(r_\s{b}-r_\s{a})(r_\s{d}-r_\s{0})}},\,k\right).
\end{split}
\label{e2.9}
\end{equation}

\noindent Using eq.\,\eqref{e1.2} and combining eqs.\,\eqref{e2.4} and \eqref{e2.5},
we can express $h$ in terms of the boundary conditions as

\begin{equation}
h =r^2_\s{0}\sqrt{\dot\theta_\s{0}^2+\sin^2\theta_\s{0}\,\dot\phi_\s{0}^2},
\label{e2.10}
\end{equation}

\noindent while evaluating eq.\,\eqref{e1.4} at $r=r_\s{0}$ allows us to
calculate $\varepsilon$: 

\begin{equation}
 \varepsilon = \dot{r}^2_\s{0} - \frac{2\,GM}{r_\s{0}} 
+ \frac{h^2}{r_\s{0}^2}-\frac{r_s h^2}{r_\s{0}^3}.
\label{e2.11}
\end{equation}

From eqs.\,\eqref{e2.10} and \eqref{e2.11} it is clear that $h$ and $\varepsilon$ 
are functions only of the initial angular position $(\theta_0,\,\phi_\s{0})$ (i.e.
they depend only on the boundary conditions) and that they are conserved along fluid
lines. 

The expressions given in this Section constitute a description of the fluid lines as
a function of the relatively simple but still general boundary conditions given in
eqs.\,\eqref{e0.1}-\eqref{e0.4}. What one needs to do to use these expressions in
practice is as follows: from the boundary conditions find the values of
$\theta_\s{a}$, $h$ and $\varepsilon$ using eqs.\,\eqref{e2.6}, \eqref{e2.10} and
\eqref{e2.11}, respectively. This determines the coefficients of the polynomial
$\mathcal{R}(r)$, as given by eq.\,\eqref{e1.7}, and its roots can then be calculated
in the way described in the Appendix. The next step is to find the numerical values
of $k$, $\varphi_\s{0}$, $\varphi_\s{a}$ and $\phi_\s{a}$ and finally one can make
use of eqs.\,\eqref{e1.10}, \eqref{e1.11} and \eqref{e2.2} to track a given fluid
element in its descent from $r_\s{0}$ towards the equatorial plane, i.e., as $\theta$
sweeps from $\theta_\s{0}$ to $\pi/2$. 

\setcounter{equation}{0}
\section{Velocity field}
\label{velocity}

  Using eqs.\,\eqref{e1.2}-\eqref{e1.4} together with eqs.\,\eqref{e2.4} and 
\eqref{e2.5}, one obtains the following expressions for the velocity field

\begin{align}
& u^t = \frac{E}{c^2}\left(1-\frac{r_s}{r} \right)^{-1},\\
& u^r = -\sqrt{\varepsilon +  \frac{2\, GM}{r} 
-\frac{h^2}{r^2}\left(1 - \frac{r_s}{r}\right)},
\label{e3.1} \\
& u^\theta = \pm\frac{h\sqrt{\cos^2\theta_\s{a}-\cos^2\theta}}{r^2 \sin\theta},
\label{e3.2}\\
& u^\phi= \frac{h\,\sin\theta_\s{a} }{r^2 \sin^2 \theta},
\label{e3.3} 
\end{align}

\noindent where the sign in eq.~(4.3) is positive for $0<\theta<\pi/2$ and
negative for  $\pi/2 < \theta < \pi$. 

It is convenient to introduce as well the velocity field as described by a set
of local observers, since their measurements correspond to physical intervals of
proper distance and proper time. Associated with each local observer one has an
orthogonal tetrad of 4-vectors constituting a local Minkowskian coordinate set
of basis vectors. Let us denote with a bar, coordinates in the local frame of
reference, i.e. let $u^{\bar\mu}$ be the components of the 4-velocity of a fluid
element passing by a local observer:

\begin{equation}
 u^{\bar\mu}=\left(\gamma\,c,\,\gamma\,\vec{V}\right),
\label{e3.4} 
\end{equation}

\noindent where $\gamma$ is the general relativistic Lorentz factor between the 
local observer and the fluid element, defined as

\begin{equation}
\begin{split}
 \gamma & =\left( 1 - \frac{\vec{V} \cdot \vec{V}}{c^2} \right)^{\text{\tiny
$-\frac{1}{2}$}} \\
	& =\frac{E}{c^2}\left(1-\frac{r_s}{r} \right)^\s{-\frac{1}{2}}
\end{split}
\label{e3.5} 
\end{equation}

\noindent and where

\begin{align}
& V^{\bar{r}} = \frac{c^2u^r}{E}, \label{e3.6}\\ 
& V^{\bar{\theta}} = \frac{r\,u^\theta}{\gamma}, \label{e3.7}\\
& V^{\bar{\phi}} = \frac{r\,\sin\theta\,u^\phi}{\gamma},\label{e3.8}
\end{align}

\noindent are the components of the three-velocity $\vec{V}$.

\setcounter{equation}{0}
\section{Density field}
\label{density}

\begin{figure}
\begin{center}
  \includegraphics[scale=0.5]{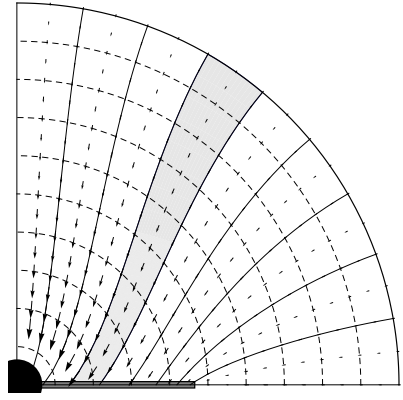}
\end{center}
\caption{Schematic illustration of the construction of a regular grid in 
$(\theta_\s{0},\,r)$ for calculating the density field. For clarity, the 
azimuthal direction is not included. The highlighted region constitutes one of 
the streamline tubes involved in the derivation of eq.\,\eqref{e4.6}.}
 \label{f2}
\end{figure}

The expressions for the streamlines and the velocity field given in Sections 
\ref{streamlines} and \ref{velocity} are independent of the value of the 
density at the boundary and hence the scale for the density can be set 
arbitrarily. In this Section we derive a numerical scheme for calculating the 
density field based on the continuity equation. For a general curved spacetime, 
this equation can be written as

\begin{equation}
  \nabla_\s{\mu} (n\,u^\s{\mu}) = \frac{1}{\sqrt{-g} } 
\frac{\partial }{\partial x^\s{\mu} } \left(\sqrt{-g}\, n\, u^\s{\mu} \right) 
    = 0,
\label{e4.1}
\end{equation}
 where $g=\text{det} [ g_\s{\mu\nu}]$ is the metric determinant, given in the 
Schwarzschild case by $g =-r^4\sin^2\theta$. Using the stationarity condition, 
eq.\,\eqref{e4.1} reduces to

\begin{equation}
   \frac{1 }{r^2 \sin \theta } \frac{\partial }{\partial x^i }( r^2 
     \sin \theta\, n\, u^i ) = 0,
\label{e4.2}
\end{equation}
 which is simply saying that the spatial divergence of the particle number flux 
3-vector \( n u^i \) is zero.  We can integrate this equation over a streamline 
tube, i.e., a volume element consisting of a collection of streamlines coming 
from an area element $\ud a|_{r_\s{0}}$ at the shell $r_\s{0}$ and ending up at 
a second sphere with arbitrary radius $ r < r_\s{0} $. Such a streamline tube 
is illustrated in Figure~\ref{f2}. By means of Gauss's theorem it follows that

\begin{equation}
  n\, u^i \ud a_i \big |_{r_\s{0}} = 
  n\, u^i \ud a_i \big |_r.
\label{e4.3}
\end{equation}

 The differential area element orthogonal to the radial direction is given by

\begin{equation}
  \ud a_r = r^2 \sin \theta\,\ud \theta\,\ud \phi,
\label{e4.4}
\end{equation}

\noindent and so eq.\,\eqref{e4.3} can be written as

\begin{equation}
n_\s{0}\,u^r_\s{0}\,r^2_\s{0}\sin \theta_\s{0}\,\ud 
\theta_\s{0} \,\ud\phi_\s{0} =
n \,u^r \,r^2 \sin \theta \,\ud \theta \,\ud \phi.
\label{e4.5}
\end{equation}
  Using the relation $\ud \theta \ud \phi=J\ud \theta_\s{0} \ud\phi_\s{0}$, 
where the Jacobian $J$ was given in eq.\,\eqref{e0.8}, we get the required 
result for the density field

\begin{equation}
 n = \frac{n_\s{0}\,u^r_\s{0}\,r^2_\s{0}\,\sin \theta_\s{0}}
{u^r\, r^2\,\sin \theta\,J }.
\label{e4.6}
\end{equation}

As long as $J>0$, which means that no intersections of the streamlines occur, and
$u^r<0$,  which implies that no turning points in the radial motion exist up
until the equator is reached, the expression for the density field given by
eq.\,\eqref{e4.6} is well defined and has no singularities.

The partial derivatives required in the calculation of $J$ represent a complex 
computation involving derivatives of an elliptic integral with respect its 
argument, modulus and integration limit. On the other hand, it is
straightforward to evaluate it numerically and so it does not seem worth
searching further for a full analytic expression.

We construct a suitable grid for calculating $J$ in the following way. We start 
with a homogeneous partition of the initial angles $(\theta_\s{0},\,
\phi_\s{0})$ and then follow the fluid lines down to the equator in regular 
radial steps. At every grid point $(\theta_\s{0},\,\phi_\s{0},\,r)$ we store 
the values of $\theta$, $\phi$ and $u^r$ and then calculate $J$ by means of 
standard finite differences. Figure~\ref{f2} illustrates the construction of such a
grid.

\setcounter{equation}{0}
\section{Example model}
\label{example}

Here we illustrate the accretion model presented in the previous sections by 
applying it with the boundary conditions considered in Paper I, i.e. ones for 
a uniform shell of matter in uniform rotation:

\begin{gather}
\varrho_\s{0} = \text{const},\label{e5.1}\\
\dot{r}_\s{0} = \text{const},\label{e5.2}\\
\dot{\phi}_\s{0} =\text{const},\label{e5.3}\\
\dot{\theta}_\s{0} = 0. \label{e5.4}
\end{gather}
 This choice leads to several simplifications, the most important being that 
the accretion flow is then axisymmetric and so, the Jacobian of the angular 
transformation in eq.\,\eqref{e0.8} simplifies to

\begin{equation}
 J = \frac{\partial \theta}{\partial \theta_\s{0}}.
\label{e5.5}
\end{equation}
 From eq.\,\eqref{e0.6}, we have that the total accretion rate is given by

\begin{equation}
\dot M = 4\pi r_\s{0}^2 \varrho_\s{0}\,|\dot r_\s{0}|,
\label{e5.6}
\end{equation}
 and, by substituting the boundary conditions in 
eqs.\,\eqref{e5.1}-\eqref{e5.4} into eq.\,\eqref{e2.10}, we get the following 
distribution of specific angular momentum:

\begin{equation}
h(\theta_\s{0}) =  h_\s{e} \sin\theta_\s{0},
\label{e5.7}
\end{equation}

\noindent where $h_\s{e}=r^2_\s{0}\,\dot\phi_\s{0}$ is the maximum value of the 
specific angular momentum, which is reached at the equator of the shell. Since 
$\dot{\theta}_\s{0} = 0$, we have that for every streamline $\theta_\s{a} = 
\theta_\s{0}$, $\varphi_\s{a} = \varphi_\s{0}$ and $\phi_\s{a} = \phi_\s{0}$ 
giving a simplification of eqs.\,\eqref{e2.2}-\eqref{e2.5}. The velocity field 
is again described by the expressions given in Section \ref{velocity} after 
making the substitution $\theta_\s{a} = \theta_\s{0}$. Regarding the density 
field, from eqs.\,\eqref{e4.6} and \eqref{e5.5}, we find the expression

\begin{equation}
 n = \frac{n_\s{0}\,u^r_\s{0}\,r^2_\s{0}\,\sin \theta_\s{0}}
{u^r\, r^2\,\sin \theta }
\left(\frac{\partial \theta}{\partial \theta_\s{0}}\right)^{-1}.
\label{e5.8}
\end{equation}

\begin{figure}
\begin{center}
  \includegraphics[scale=0.6]{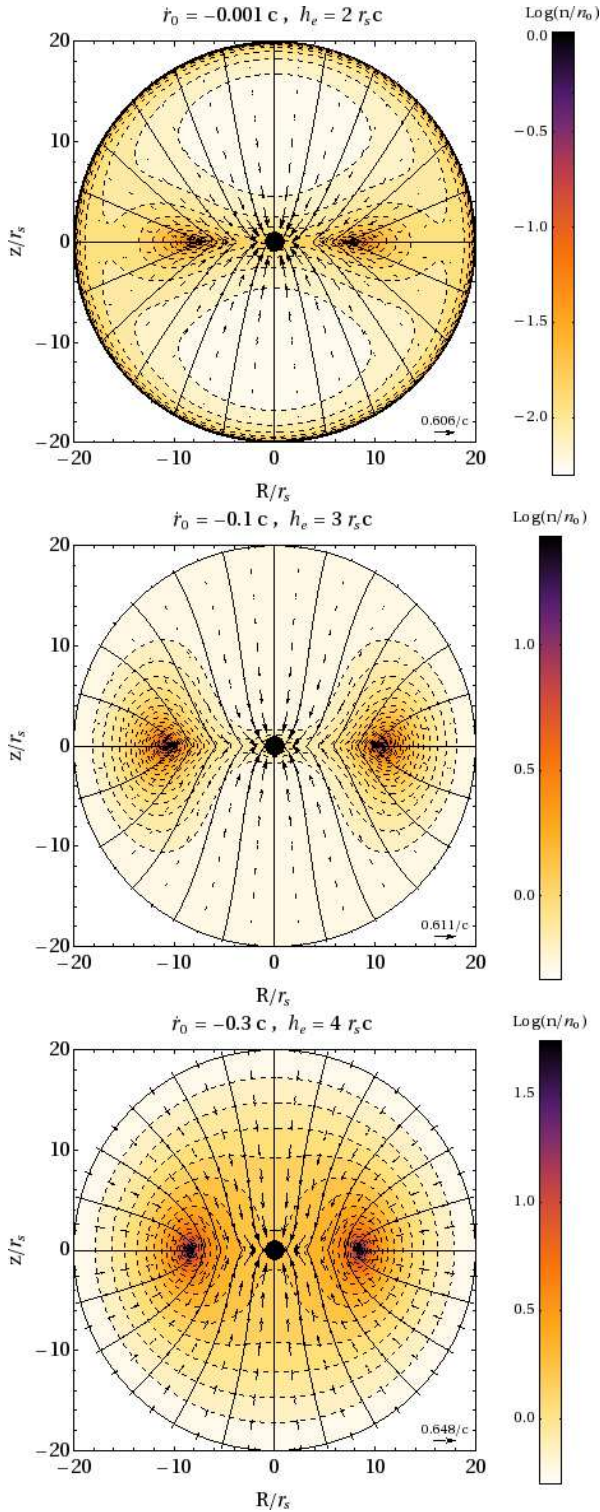}
\end{center}
 \caption{Streamlines, velocity field and density contours for different flow 
parameters. In each of the panels we have taken $r_\s{0} = 20\,r_s$, while the 
values of the remaining parameters are indicated in each plot. The plots are 
projections onto the $R$-$z$ plane and the colour coding corresponds to the 
logarithm of the particle number density, Log$(n/n_\s{0})$. The arrows 
correspond to the $V^{\bar{r}}$ and $V^{\bar{\theta}}$ components of the 
velocity field.}
 \label{f3}
\end{figure}

Figure~\ref{f3} shows the projection onto the $R$-$z$ plane ($R=r\sin\theta$, 
$z=r\cos\theta$) of the streamlines, velocity field and density contours for 
three different combinations of $\dot r_\s{0}$ and $h_\s{e}$. In each case we 
have set $r_\s{0} = 20\, r_s$.

\begin{figure}
\begin{center}
  \includegraphics[scale=0.6]{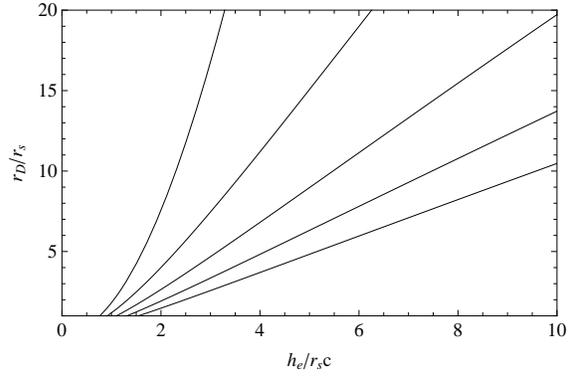}
\end{center}
 \caption{The radius of the outer edge of the disc formed as matter reaches the 
equatorial plane, $r_\s{D}$, is plotted against the specific angular momentum 
at the equator, $h_\s{e}$. A fixed value of $r_\s{0}= 20\,r_s$ is considered. 
From top to bottom, each curve corresponds to a different inward radial 
velocity with  values $|\dot{r}_\s{0}|/c=0,\, 0.2,\, 0.4,\, 0.6,\,0.8$ respectively.}
 \label{f4}
\end{figure}

 For the choice of boundary conditions being used here, the radius of the outer 
edge of the disc formed as matter reaches the equatorial plane, $r_\s{D}$, can 
be calculated from eq.\,\eqref{e1.11} and \eqref{e2.2}, taking first 
$\theta=\pi/2$ and then $\theta_\s{0}=\pi/2$, giving

\begin{equation}
\xi_\s{D} = \frac{\sqrt{\varepsilon(r_{\s{a}}-r_{\s{c}})
(r_{\s{d}}-r_{\s{b}}) } }{2\,h_\s{e}}\left(\varphi_\s{0}+\frac{\pi}{2}\right),
\label{e5.9}
\end{equation}

\noindent and then substituting the result into eq.\,\eqref{e1.10}. In 
Figure~\ref{f4}, we have plotted $r_\s{D}$ as a function of $h_\s{e}$ as 
obtained from eq.\,\eqref{e5.9} for a fixed value of $r_\s{0}$ and five 
different values of $\dot{r}_\s{0}$. Note that $r_\s{D}$ is the radius at 
which the particle with the largest angular momentum {\it first} impacts on the
equatorial plane; no further motion is then considered. For sufficiently
small $r_\s{D}$, the ``disc'' will be one within which accretion into
the black hole then proceeds on a free-fall time scale, as discussed earlier,
and only for larger values of $r_\s{D}$ could it give rise to a Keplerian-type
disc. Naturally, $r_\s{D}$ increases monotonically with increasing $h_\s{e}$,
while it decreases with increasing $|\dot{r}_\s{0}|$. For a given specific
angular momentum, $r_\s{D}$ can be substantially smaller when larger values are
taken for $|\dot{r}_\s{0}|$. 


\begin{figure*}
\begin{center}
\includegraphics[scale=0.56]{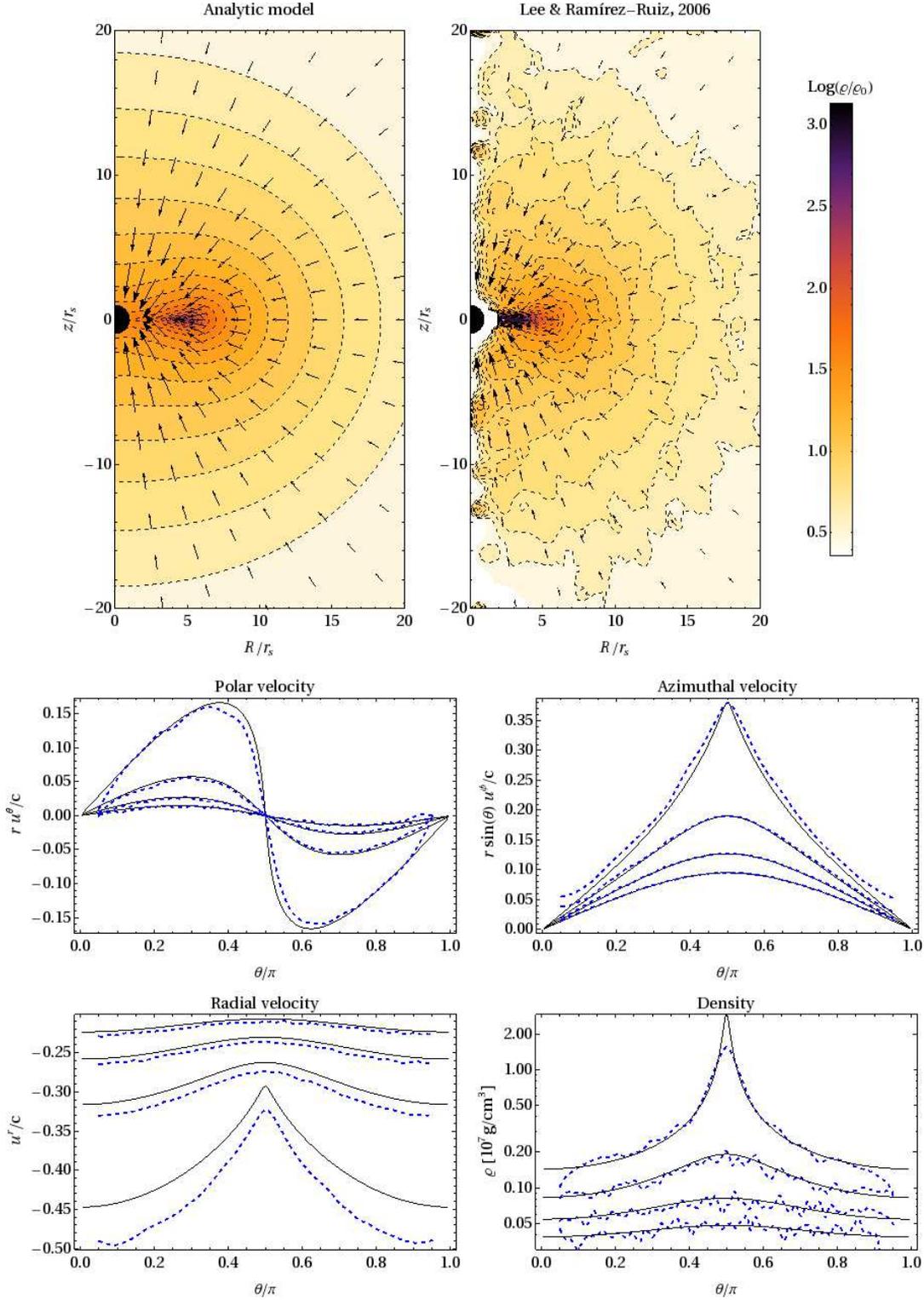}
\end{center}
 \caption{Comparison between the analytic model and one of the LR06 
simulations. The plots are for an accretion flow towards a black hole with mass 
$M=4M_\odot$, starting from a spherical shell at $r_\s{0} = 50\,r_s$ where the 
matter density and radial inward velocity are uniform and given by 
$\varrho_\s{0} = 5.29\times 10^6\, \text{g}/\text{cm}^3$ and $\dot r_\s{0} = 
-\sqrt{1/50}\,c = - 0.14\,c$, respectively. The specific angular momentum 
distribution at the shell corresponds to uniform rotation with a maximum of 
$h_\s{e} = 1.9\, r_sc$. With these boundary conditions, the total accretion rate
is $\dot M = 0.01 M_\odot / s$. The top panels show a projection of the
accretion flow onto the $R$-$z$ plane, together with isodensity contours of the
analytic solution (left) and the LR06 numerical simulation (right). The
remaining four panels show the velocity components and the density at the radial
cuts $r/r_s = 20,\,15,\,10,\,5$ with the analytic and numerical results being
represented by solid and dashed lines respectively.}
 \label{f5}
\end{figure*}

Let us now consider boundary conditions corresponding to the collapse of a 
massive stellar core such as those studied numerically by LR06. In that work, 
the authors investigated the formation of an inviscid, small-scale accreting 
disc by making a 2D Smoothed Particle Hydrodynamics (SPH) simulation with 
relativistic effects being mimicked with a \cite{paczyski80} (PW) 
pseudo-Newtonian potential. They included a detailed treatment of neutrino 
cooling and considered an equation of state with contributions from radiation, 
$e^\pm$ pairs, $\alpha$ particles and free nucleons. We have made comparisons 
for several of the models discussed in LR06, finding similar results in all 
cases. For illustration, in Figure~\ref{f5} we show just one of them, namely 
the numerical run in which LR06 took a central black hole of $4M_\odot$ and an
external spherical boundary at $r_\s{0} = 50\,r_s$ from which SPH particles were
continuously injected with a constant accretion rate of $\dot M = 0.01 M_\odot
/\text{s}$. As radial infall velocity, they took the velocity of free fall from
infinity, i.e. $\dot r_\s{0} = -\sqrt{1/50}\,c$, and the specific angular
momentum of the fluid elements at $r_\s{0}$ was assumed to follow a rigid body
rotation distribution with a maximum of $h_\s{e} = 1.9\,r_sc$ at the equator of
the shell.

The top left panel in Figure~\ref{f5} shows the accretion flow as calculated 
from the analytic model, while the top right panel shows a late-time snapshot 
of the LR06 simulation, taken when a quasi-stationary state had been reached. 
In general, there is quite good agreement between them until the streamlines 
have approached the equatorial disc. This is not surprising, bearing in mind 
that the flow is highly supersonic all the way down to the vicinity of the 
equatorial disc, where strong shocks then appear in the hydrodynamical 
simulation because of collisions between streamlines coming from opposite 
points in the two hemispheres. The other four panels present a detailed 
comparison of the spatial components of the velocity and the density at four 
spherical cuts. Here, we see very good agreement between the analytical and 
numerical results for $u^{\theta}$ and $u^{\phi}$. For $u^r$ and $\varrho$,
there is quite good qualitative agreement, although the numerical results for
$\varrho$ suffer from numerical noise inherent in the interpolation scheme at
low particle number densities, and the numerical results for $u^r$ show higher
radial infall velocities, as expected given the use of the PW pseudo-Newtonian
potential there, which artificially enhances the radial acceleration.

\setcounter{equation}{0}
\section{Relativistic extension of Ulrich's model}
\label{ulrich}

  In this Section we adopt boundary conditions corresponding to the models of 
\cite{ulrich}, \cite{belo01} and HM07, i.e. we consider infall from an 
infinitely large spherical shell of matter with all of the fluid elements 
having parabolic-like energies ($r_\s{0} \rightarrow\infty$, $\dot r_\s{0} = 0$ 
and hence $\varepsilon =0$).

In Ulrich's model, the location of the outer edge of the equatorial disc is 
simply related to $h_\s{e}$ through 

\begin{equation}
 r_\s{k} = \frac{h_\s{e}^2}{GM}. \label{e6.1}
\end{equation}

\noindent In Newtonian gravity, $r_\s{k}$ corresponds to the Keplerian radius
of a circular orbit with specific angular momentum $h_\s{e}$ as well as to the
semi-latus rectum of a parabolic orbit with this specific angular momentum.

Note that a vanishing initial radial velocity implies $r_\s{b} = r_\s{0} $
and, consequently, $r_\s{c} = 0$ while 

\begin{equation}
 r_\s{a,d} = \frac{h^2}{4\,GM}\left[ 1 \pm
\sqrt{1-4\left(\frac{r_sc}{h}\right)^2 } \right]. 
\label{e6.2}
\end{equation}

Taking the limit $r_\s{0} \rightarrow\infty$ in eq.\,\eqref{e1.10}-\eqref{e1.12}
and noticing that $\varepsilon\, r_\s{0} \rightarrow -2\,GM $, one gets the
following expression for the streamlines

\begin{equation}
 r = \frac{ r_\s{a} -r_\s{d}\,\sn^2(\xi,\,k) }{ \cn^2(\xi,\,k) },
\label{e6.3}
\end{equation}

\noindent with

\begin{gather}
  \xi =  \sqrt{\frac{GMr_\s{a}}{2}}\,\frac{\varphi}{h},\label{e6.4}
\\
    k = \sqrt{\frac{r_\s{d} }{r_\s{a} }}.\label{e6.5}
\end{gather}

\begin{figure}
\begin{center}
\includegraphics[scale=0.75]{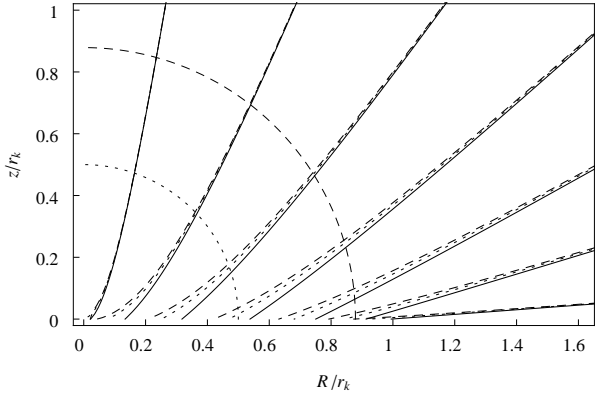}
\end{center}
 \caption{Streamlines for the relativistic extension of Ulrich's model 
($r_\s{0}\rightarrow\infty$, $\dot r_\s{0}=0$). The continuous line corresponds 
to $h_\s{e}\gg r_sc$, i.e. to the non-relativistic limit in which Ulrich's 
model is recovered. The dotted line corresponds to $h_\s{e}=r_sc$  while the 
dashed line is for $h_\s{e}=h_c\approx0.754\,r_sc$, in which case the disc
forms entirely inside $r_s$, i.e. $r_\s{D} = r_s$. The quarter-circles
correspond the location of the event horizon in each case.}
 \label{f6}
\end{figure}

  Figure~\ref{f6} shows plots of the streamlines for different values of 
$h_\s{e}$. From these, we see how the radius of the equatorial disc decreases 
as $h_\s{e}$ decreases starting from $r_\s{D}=r_\s{k}$ for $h_\s{e}\gg r_sc$ 
(corresponding to the disc radius in the Ulrich model) down to $r_\s{D} = r_s$ 
when $h_\s{e}=h_\s{c}\approx0.754\,r_sc$.

HM07 followed an analytic approach similar to the one presented here, but used 
an incorrect transformation for the polar and azimuthal angles between the 
systems of reference $\mathcal{O}$ and $\mathcal{O'}$. In particular, eq.\,(27) 
in HM07 is not correct and should be substituted by eqs.\,\eqref{e2.2} and 
\eqref{e2.3} of the present paper.

\setcounter{equation}{0}
\section{Non-relativistic limit}
\label{newton}

In this Section we consider the non-relativistic limit, for which $h_\s{e} \gg
r_sc$. By taking this limit in eq.\,\eqref{e2.11}, it follows that 
$\varepsilon$ becomes equal to twice the total Newtonian specific energy:

\begin{equation}
 \varepsilon = \dot{r}^2_\s{0} + \frac{h^2}{r_\s{0}^2} - \frac{2\,GM}{r_\s{0}},
\label{e7.1}
\end{equation}
 and, from eq.\,\eqref{e1.7}, one has that $r = 0$ is a double root of 
$\mathcal{R}(r)$ (i.e., $r_\s{c}=r_\s{d}=0$), with the remaining two roots 
being real and given by

\begin{equation}
  r_\s{a,b} = \frac{GM }{\varepsilon } \left( -1 \pm e \right),
\label{e7.2}
\end{equation}
 where $e = \sqrt{1 + \varepsilon (h/GM)^2}$.

  From eq.\,\eqref{e1.12} it follows that $k=0$. For a null value of the 
modulus, the Jacobi elliptic functions reduce to ordinary trigonometric 
functions:

\begin{equation}
  \cn(x,\,0)=\cos(x),\qquad\sn(x,\,0)=\sin(x),
\label{e7.5}
\end{equation}
 and so eqs.\,\eqref{e1.10} and \eqref{e1.11} become

\begin{equation}
\begin{split}
r &=\frac{r_\s{a}\,r_\s{b}}{r_\s{a} + (r_\s{b} - r_\s{a}) \cos^2 (\varphi/2)}\\
  &=\frac{h^2/GM}{1 + e \cos\varphi }.
\end{split}
\label{e7.6}
\end{equation}

\noindent The second equality in eq.\,\eqref{e7.6} is the well-known expression for 
motion along a conic section with eccentricity $e$, representing all of the possible
types of orbit in Newtonian gravity.

 All of the expressions derived in Section \ref{angles} concerning the angles 
$\varphi$, $\varphi_\s{0}$, $\varphi_\s{a}$, $\theta$, $\theta_\s{0}$ and
$\theta_\s{a}$ remain valid, but note that the expression for the initial orbital
angle is simplified:

\begin{equation}
\varphi_\s{0}=\cos^{-1}\left(\frac{h^2/GM-r_\s{0}}{e\,r_\s{0}}\right).
\label{e7.8}
\end{equation}

The velocity field is given by eqs.\,\eqref{e3.1}-\eqref{e3.3} after taking $r_s=0$
and noticing that, within the non-relativistic limit, proper time intervals become
identical with coordinate time intervals: $\ud \tau = \ud t$. The Newtonian results
presented in Paper I can be recovered from the expressions given here once the
boundary conditions in Section \ref{example} have been adopted.\footnote{Note that in
Paper I, the azimuthal angle was measured starting from the apastron instead of from
the periastron as we do here.} The density field is again given by eq.\,\eqref{e5.8}.
Note that for the particular boundary conditions considered in Paper I, direct
evaluation of $J$ is straightforward and gives the analytic expression presented
there.

\setcounter{equation}{0}
\section{Summary and Discussion}
\label{discussion}

In this paper we have presented an analytic solution for the streamlines of 
pressureless matter being steadily accreted towards a Schwarzschild black hole. 
The accretion flow is taken to start from a spherical surface far away from the 
central mass, with a wide range of boundary conditions being used. The fluid 
streamlines are then tracked down to the point at which they either become 
incorporated into a thin equatorial disc or pass inside the black hole event 
horizon. We have presented a simple numerical algorithm for calculating the 
density field.

In our model, the fluid streamlines correspond to timelike geodesics of 
Schwarzschild spacetime. Using Jacobi elliptic functions and some standard 
identities, we have developed a novel approach for describing all of the 
different types of trajectory with a single analytical expression.

This work extends the Newtonian model presented in Paper I to a general 
relativistic one in Schwarzschild spacetime and constitutes the analytic
solution corresponding to the scenario studied numerically by \cite{belo01}.
We are currently working on the further generalisation of the model to Kerr
spacetime. For appropriate boundary conditions, the present solution generalises
the classical \cite{ulrich} model and gives corrected expressions for the HM07
model. We have shown that our model recovers the well-known Newtonian
expressions in the non-relativistic limit.

Our analytic solution can be used as a benchmark for testing general
relativistic hydrodynamical codes. Clearly, such codes should be able to 
correctly reproduce geodesic motion for a cloud of non-interacting particles in 
a fixed metric.

Despite the fact that the present model leaves out many physical processes 
which are relevant for the study of realistic accretion onto black holes, it 
does include the two main factors determining the bulk dynamics: gravity and 
rotation. Furthermore, given its flexibility for setting boundary conditions, 
it can be useful as a computationally inexpensive and efficient tool for 
exploring the parameter space in applications where the ballistic and steady 
state hypothesis are approximately valid, such as: sub-Eddington accretion 
towards a supermassive black hole in a galactic nucleus, wind-fed and 
Roche-lobe fed X-ray binaries, and collapsar GRB progenitors. In this way one 
can gain physical insight and get order-of-magnitude estimates of the energy 
budget before undertaking full-scale simulations. Having such an exploratory 
tool can be especially relevant if one bears in mind that the parameter domain
for many of these systems is vast and often uncertain. An example is the 
comparison made in Section \ref{example} with the LR06 SPH simulation, where 
rather good agreement was found between the analytic model and the numerical 
results.

\section{Acknowledgements}

The authors thank Stephan Rosswog for insightful discussions, and are grateful
to William Lee for his valuable comments and for providing the LR06 numerical
results against which the analytical solution of this paper was compared. ET
acknowledges Jorge Moreno for helpful comments on an earlier version of the
manuscript. This work was partially supported by a DGAPA-UNAM grant(PAPIIT
IN116210-3). SM acknowledges financial support from CONACyT 26344.

\bibliographystyle{mn2e}
\bibliography{acc}

\appendix

\renewcommand{\thesection}{}
\renewcommand{\theequation}{A.\arabic{equation}}

\section{Solution of the radial equation}
\label{appendix}

In this appendix we discuss the solution for the radial motion in more
detail. Explicit expressions for the roots of the fourth degree polynomial
$\mathcal{R}(r)$ defined in eq.\,\eqref{e1.7} are easily given in terms 
of the following auxiliary quantities \citep[see e.g.][]{abramowitz}

\begin{gather}
    Q = (2GM)^2 + 3\,\varepsilon h^2, \label{a2}\\
    R = (2GM)^3 + 9\,\varepsilon h^2 
\left(GM + \frac{3}{2} r_s\,\varepsilon \right), \label{a3}\\
    D^2 = R^2 - Q^3.
\label{a4}
\end{gather}

\noindent When $D^2<0$, all of the roots are real and are given by 

\begin{gather}
r_\s{1} = 0\\
r_\s{2} = \frac{ 2 }{ 3\, \varepsilon } \left[ \sqrt{ Q } 
   \cos\left(  \Psi - \frac{\pi}{3} \right)  - GM \right] , \label{a5} \\
r_\s{3} = \frac{ 2 }{ 3\, \varepsilon } \left[  \sqrt{ Q } 
   \cos\left(  \Psi + \frac{\pi}{3} \right) - GM \right] ,  \label{a6} \\
r_\s{4} = \frac{ 2 }{ 3\, \varepsilon } \left[  \sqrt{ Q } 
  \cos( \Psi + \pi ) - GM \right] ,
\label{a7}
\end{gather}

\noindent with $\Psi$ being defined through 

\begin{equation}
  \cos (3\, \Psi) = \frac{ R }{ Q^{3/2} }.
\label{a8}
\end{equation}

\noindent When $r_\s{4}<0$, we interchange $r_\s{2}$ and $r_\s{3}$ in order to
satisfy $r_\s{2}<r_\s{3}$. In this way, when real, the four roots are ordered as

\begin{equation}
  r_\s{1}<r_\s{2}<r_\s{3}<|r_\s{4}|.
\label{a9}
\end{equation}

  On the other hand, for $D^2>0$ two of the roots form a complex conjugate 
pair. Making the definitions

\begin{gather}
S = \sqrt[3]{ D - R }, \label{a10}\\ 
T = \sqrt[3]{ D + R }, \label{a11}
\end{gather}

\noindent the non-zero roots in this case are given by

\begin{gather}
  r_\s{2,3} =  \frac{ 1 }{ 6\, \varepsilon }\left[  T - S - 4GM  
             \pm i \sqrt{3}(S+T)\right] , \label{a12}\\
  r_\s{4} = \frac{ 1 }{ 3\, \varepsilon }(S - T - 2GM),
\label{a13}
\end{gather}

%

Since the radial motion is constrained to satisfy $\mathcal{R}(r)>0$ 
then, in the $D<0$ case, the radial coordinate is bounded as

\begin{equation}
r \le r_\s{2}, \quad \text{or} \quad r_\s{3} \le r.
\label{a14}
\end{equation}

\noindent If the first inequality holds, we write

\begin{equation}
 r_\s{a} = r_\s{1},\ r_\s{b} = r_\s{2},\
 r_\s{c} = r_\s{3},\ r_\s{d} = r_\s{4},
\label{a15}
\end{equation}

\noindent but, if the second inequality holds, we write

\begin{equation}
 r_\s{a} = r_\s{3},\ r_\s{b} = r_\s{4},\ 
 r_\s{c} = r_\s{1},\ r_\s{d} = r_\s{2}.
\label{a16}
\end{equation}
 
\noindent In this way the solution to eq.\,\eqref{e1.9} is always given by
eq.\,\eqref{e1.10}.

On the other hand, in the case $D>0$, we take

 \begin{equation}
 r_\s{a} = r_\s{1},\ r_\s{b} = r_\s{2},\ 
 r_\s{c} = r_\s{3},\ r_\s{d} = r_\s{4},
\label{a17}
\end{equation}

\noindent and once again the expression in eq.\,\eqref{e1.10} is a formal 
solution to eq.\,\eqref{e1.9}. However, direct evaluation of this involves the 
use of complex quantities as intermediate steps. It is possible to rewrite 
eq.\,\eqref{e1.10} as an expression involving just real quantities. For doing 
this, we introduce the following two real constants (having in mind the
fact that $r_\s{1}=0$)

\begin{gather}
\alpha = \pm\sqrt{(r_\s{4} - r_\s{2})(r_\s{4} - r_\s{3})},\label{a18} \\
\beta = \sqrt{r_\s{2}r_\s{3}},
\label{a19}
\end{gather}

\noindent where the sign of $\alpha$ coincides with that of $\varepsilon$.
We then define

\begin{gather}
 \xi_\s{2} =  \frac{\sqrt{ \varepsilon\alpha\beta }}{h}\, \varphi 
= 2\sqrt{\mp k}\, \xi,  \label{a20}\\
 k^2_\s{2} =  \frac{( \alpha + \beta )^2 - r_\s{4}^2 }{ 4\, \alpha\beta }  
=  \frac{(1\mp k)^2}{\mp 4\, k},
\label{a21}
\end{gather}

\noindent where the sign accompanying $k$ in last equations is the opposite of
the one for $ \alpha$ . If we now invok the following identity for Jacobi
elliptic functions \citep[see, e.g.][]{hancock}

\begin{equation}
 \mp k\,\sn^2\left(\xi,\, k\right)= \frac{1-
\cn\left(\xi_\s{2},\,k_\s{2}\right)}
 {1+ \cn\left(\xi_\s{2},\,k_\s{2}\right)},
\label{a22}
\end{equation}

\noindent we can rewrite equation~\eqref{e1.10} as

\begin{equation}
 r = \frac{ \beta\, r_\s{4}\left[1-\cn \left(\xi_\s{2},\,k_\s{2}\right)\right]}
 { \beta-\alpha - ( \alpha +\beta) \cn \left(\xi_\s{2},\,k_\s{2}\right) }.
\label{a23}
\end{equation}

\noindent Note that, in this case,

\begin{equation}
 \varphi_\s{0} = \frac{- h }{ \sqrt{ \varepsilon\alpha\beta } }\,
\cn^{-1} \left[\frac{ \beta\,r_\s{4} + (\alpha-\beta)r_\s{0}  }
{ \beta\,r_\s{4} - (\alpha+\beta)r_\s{0} },\,k_\s{2}\right] .
\label{a24}
\end{equation}

In summary, the expression for the trajectory of a timelike geodesic in
Schwarzschild spacetime can be written as follows:

\begin{equation}
 r =\left\{\begin{array}{l} 
{\rm (i) \ }\frac{\displaystyle r_\s{2}\, r_\s{4}\,\sn^2 \left(\xi,\,k\right) }
{\displaystyle r_\s{4} - r_\s{2}\,\cn^2 \left(\xi,\,k\right) }\\[15pt]
{\rm (ii) \ }\frac{\displaystyle r_\s{4} (r_\s{3}-r_\s{2})+r_\s{2}(r_\s{4}-r_\s{3})
\cn^2 \left(\xi,\,k\right) }
{\displaystyle r_\s{3}-r_\s{2}+(r_\s{4}-r_\s{3})\cn^2\left(\xi,\,k\right) }
\\[15pt]
{\rm (iii) \ }\frac{\displaystyle\beta\,r_\s{4}\left[1-\cn\left(\xi_\s{2},\,k_\s{2}
\right)\right]}
 {\displaystyle \beta-\alpha - ( \alpha +\beta)
\cn\left(\xi_\s{2},\,k_\s{2}\right) }
\end{array}\right.,
\label{a25}
\end{equation}

\begin{equation}
\xi = \frac{\sqrt{\varepsilon\,r_\s{3}(r_\s{2}-r_\s{4})}}{2\,h}\,\varphi, \qquad
k  = \sqrt{\frac{r_\s{2}(r_\s{4}-r_\s{3})}{r_\s{3}(r_\s{4}-r_\s{2})}},
\end{equation}

\noindent where (i) applies if there are 4 real roots and $r<r_\s{2}$, (ii) is 
used for 4 real roots but $r>r_\s{3}$ and (iii) applies when there are just two 
real roots.

\label{lastpage}

\end{document}